\documentclass[submission]{eptcs}
\providecommand{\event}{GRAPHITE 2012}
\usepackage{prelude}

\newcommand{\GROOVE} {{\sc groove}\xspace}

\newcommand{\Lab}{\mathsf{Lab}\xspace}
\newcommand{\LabU}{\mathsf{\Lab^{U}}}
\newcommand{\LabB}{\mathsf{\Lab^{B}}}
\newcommand{\EB}{E^\mathsf{B}}

\newcommand{\lab}{\mathsf{lab}}
\newcommand{\ine}{\mathsf{in}}
\newcommand{\oute}{\mathsf{out}}

\newcommand{\iso}{\sim}

\newcommand{\isomap}{\varphi}

\newcommand{\transforms}[1]{\stackrel{#1}\longrightarrow}

\newcommand{\neigheq}{\equiv}
\newcommand{\seqclassz}[1]{{#1\!\mathop{/}\!\!\neigheq_0}}
\newcommand{\seqclasseq}[1]{{#1\!\mathop{/}\!\!\simeq}}
\newcommand{\eqclass}[2]{[#1]_{#2}}

\newcommand{\mult}{\mathsf{mult}}
\newcommand{\out}{\mathsf{o}}
\newcommand{\inc}{\mathsf{i}}
\newcommand{\node}{\mathsf{n}}
\newcommand{\multn}{\mult^{\node}}
\newcommand{\multi}{\mult^{\inc}}
\newcommand{\multo}{\mult^{\out}}

\newcommand{\concr}{\mathsf{concr}}

\newcommand{\calM}{\mathbf{M}}

\newcommand{\NatO}{\Nat^{\omega}}

\newcommand{\subsump}{\sqsubseteq}

\newcommand{\nbound}{\mathbf{b}}

\newcommand{\calMb}{\mathbf{M}^\nbound}

\newcommand{\multz}{\mathbf{0}}
\newcommand{\multone}{\mathbf{1}}
\newcommand{\multt}{\mathbf{2}}
\newcommand{\multzw}{\multz^{+}}
\newcommand{\multow}{\multone^{+}}
\newcommand{\multtw}{\multt^{+}}

\newcommand{\abstr}{\textsf{abstract}\xspace}
\newcommand{\prematch}{\textsf{prematch}\xspace}
\newcommand{\materialise}{\textsf{materialise}\xspace}
\newcommand{\apply}{\textsf{apply}\xspace}
\newcommand{\normalise}{\textsf{normalise}\xspace}

\newcommand{\isfresh}{\textsf{isFresh}\xspace}
\newcommand{\cert}{\textsf{certificate}\xspace}

\begin{document}

\title{Graph Subsumption in Abstract State Space Exploration}%
\author{%
Eduardo Zambon%
\thanks{The work of this author is supported by the GRAIL project, funded by %
NWO (Grant 612.000.632).}%
\qquad\qquad%
Arend Rensink%
\institute{Formal Methods and Tools Group\\%
Computer Science Department\\%
University of Twente, The Netherlands}%
\email{\{zambon, rensink\}@cs.utwente.nl}%
}%
\def\titlerunning{Graph Subsumption in Abstract State Space Exploration}%
\def\authorrunning{E. Zambon \& A. Rensink}%

\maketitle

\begin{abstract}
In this paper we present the extension of an existing method for abstract graph-based
state space exploration, called neighbourhood abstraction, with a reduction
technique based on subsumption. Basically, one abstract state subsumes another
when it covers more concrete states; in such a case, the subsumed state need
not be included in the state space, thus giving a reduction. We explain the
theory and especially also report on a number of experiments, which show that
subsumption indeed drastically reduces both the state space and the resources
(time and memory) needed to compute it.
\end{abstract}


\section{Introduction}
\stlabel{intro}

Traversal of the state space of systems is the cornerstone of many
verification/analysis methods, \eg, model checking~\cite{BK08}, and is therefore
a subject under intense investigation. In particular, techniques for pruning the
search space (\eg, partial-order reduction~\cite{God96}) and duplicate state
detection (\eg, collapsing under isomorphism~\cite{Ren06}) were shown to be
essential ingredients in the fight to tame the ever looming problem of state
space explosion. However, important classes of systems have infinite state
spaces and therefore cannot be (fully) explored using traditional traversal
techniques.

One way to address this problem is to perform \emph{state abstraction}, where
``similar'' concrete states are collapsed under an abstract representative, with
the behaviour of the abstract state encompassing all possible behaviour of the
collapsed concrete states. This notion of abstraction is the basis of well-known
techniques such as abstract interpretation~\cite{CC77} and shape
analysis~\cite{SRW02}.

State ``similarity'' is the point where many abstractions differ; in order to
define what ``similar'' means one has to look at the underlying framework one
uses to represent systems. In our case, we use \emph{graph transformation} as
the framework for modelling system behaviour and therefore our abstraction works
over graphs. In this context of graph transformation, many theoretical studies
on suitable abstractions have been proposed~\cite{Ren04, RD06, BBKR08, RN08,
BCK01, BW07, SWJ08}. However, only the last three of these were backed-up
by tool support.

In previous work~\cite{RZ10}, we presented a prototype implementation of the
\emph{neighbourhood abstraction} theory developed in~\cite{BBKR08}, as an
extension of the \GROOVE tool set~\cite{GROOVE,STTT}. The main
functionality of \GROOVE is the ability to explore the state space of graph
transformation systems (more details in \stref{abstraction}), but a concrete
exploration can only traverse part of the state space of an infinite state
system. The abstraction extension allows \GROOVE to generate a \emph{finite}
abstract state space that over-approximates the original concrete one.
Over-approximation guarantees soundness of the verification, \ie, properties
that hold in the abstract domain also hold in the concrete counterpart.

The main goal of the prototype implementation was to serve as a practical
proof-of-concept of the theoretical ideas. Unsurprisingly, performance was not
optimal and only a few small systems could be properly analysed. Since then, we
completely re-implemented the abstraction code and incorporated many performance
improvements. This paper describes one of such improvements, based on the
concept of \emph{state subsumption}. We illustrate the performance gain
provided by the subsumption with experiments using test cases from different
areas of computer science.

The rest of this paper is organised as follows. First, we present the key
concepts of graph transformation and abstraction in \stref{abstraction}. In
\stref{subsumption}, we introduce the subsumption relation for abstract states
and we show how it can be used during exploration. In \stref{experiments}, we
present the experiments performed and analyse the results. Related work
is discussed in \stref{relwork}. Finally, \stref{conclusion} concludes the
paper.


\section{Graph Production Systems and Abstraction}
\stlabel{abstraction}

\emph{Graph transformation}~\cite{Handbook-GG-2} is a rewriting technique that
operates over graphs. In its simplest form, a \emph{transformation rule}
consists of a left-hand side (LHS) and a right-hand side (RHS) graph and
specifies the changes that should be performed to a \emph{host graph}. Applying
a rule $r$ to a host graph $G$ basically amounts to finding a match $m$ of the
LHS of $r$ in $G$ and replacing this matched part of $G$ by the RHS of $r$, thus
producing a new graph $H$. We write $G \transforms{r,m} H$ to denote a rule
application and we write $G \transforms{r} H$ if there exists a match $m$ such
that $G \transforms{r,m} H$.

A graph production system (or graph grammar) is formed by a set of graph
transformation rules~$R$ and a start host graph. State space exploration of the
grammar consists of performing all possible applications of the rules from $R$
into the start graph, and repeating this process to all newly generated graphs.
The state space obtained in this way can be represented by a Graph Transition
System (GTS), which is a labelled transition system where states are host graphs
and transitions are rule applications, \ie, a pair of rule $r$ and associated
match $m$. Once generated, a GTS can be analysed as usual, for example by model
checking properties written as temporal logic formulae (\eg, using Computation
Tree Logic -- CTL). Clearly, if the rewrite system modelled by a graph grammar
is non-terminating, the state space is infinite and thus a GTS cannot be fully
constructed by a normal exploration method. To handle infinite state systems,
some form of abstraction is required. One of such abstractions, called
neighbourhood abstraction, is presented in \stref{neigh-abs}. First, we
formalise the graph representation that we use.

We assume the existence of a finite set of labels $\Lab$, partitioned into
unary and binary label sets, denoted $\LabU$ and $\LabB$, respectively. We work
with simple directed graphs, with labels taken from $\Lab$.

\begin{definition}[Graph]
A graph is a tuple $G = \tup{N,E}$, where $N$ is a finite set of nodes and
$E \subseteq N \times \Lab \times N$ is a finite set of directed labelled
edges, such that $\tup{v,l,w} \in E$ with $l\in \LabU$ implies $v = w$.
\defend
\end{definition}

Given $\tup{v, l, w} \in E$, $v$ and $w$ are called source and target nodes,
respectively; and $l$ is the edge label. We simulate node labels with
self-edges labelled with unary labels. Given $v \in N$, the \emph{set} of
labels of node~$v$, denoted $\lab(v)$, is defined as $\lab(v) = \set{l \in
\LabU \st \tup{v,l,v} \in E}$. For convenience, we write $\EB$ to denote the
set of edges with binary labels, \ie, $\EB = \set{\tup{v, l, w} \in E \st l \in
\LabB}$.

\subsection{Example of a Graph Grammar}

As an example we use a graph grammar modelling the behaviour of a firewall in a
network, taken from~\cite{KK06}. A \emph{firewall} has \emph{inner} and
\emph{outer} \emph{interfaces}, to which \emph{locations} can be connected.
Locations are
marked with the kind of interface they are connected to. Data are represented as
\emph{packets}, which can be transferred between locations or through the
firewall. Packets can either be \emph{safe} or \emph{unsafe}. Safe packets can
be \emph{at} any location but unsafe packets cannot exist at inner locations.
\fref{ex-host-firewall} shows an example configuration of a network with simple
abbreviations used for conciseness. The network has one outer location and
two inner ones, and there are five packets being transmitted.

\begin{figure}[t]
\centering
%
\begin{tikzpicture}[
scale=\tikzscale]
\node[node] (n9)  at (1.055, -0.555) {\ml{\textbf{L}\\\textit{i}}};
\node[node] (n2)  at (3.355, -0.855) {\ml{\textbf{IF}}};
\node[node] (n0)  at (2.505, -0.855) {\ml{\textbf{FW}}};
\node[node] (n6)  at (4.055, -0.855) {\ml{\textbf{L}\\\textit{o}}};
\node[node] (n1)  at (1.755, -0.855) {\ml{\textbf{IF}}};
\node[node] (n4)  at (1.055, -1.355) {\ml{\textbf{L}\\\textit{i}}};
\node[node] (n3)  at (4.655, -0.455) {\ml{\textbf{P}\\\textit{s}}};
\node[node] (n5)  at (5.055, -0.855) {\ml{\textbf{P}\\\textit{u}}};
\node[node] (n7)  at (4.655, -1.255) {\ml{\textbf{P}\\\textit{u}}};
\node[node] (n8)  at (0.355, -0.555) {\ml{\textbf{P}\\\textit{s}}};
\node[node] (n10)  at (0.355, -1.155) {\ml{\textbf{P}\\\textit{s}}};
\path[edge](n0.east |- 3.355, -0.855) -- node[lab]{out} (n2) ;
\path[edge](n0.west |- 1.755, -0.855) -- node[lab]{in} (n1) ;
\path[edge] (n9)  -- node[lab]{c} (n1) ;
\path[edge](n6.west |- 3.355, -0.855) -- node[lab]{c} (n2) ;
\path[edge] (n4)  -- node[lab]{c} (n1) ;
\path[edge](n9.south -| 1.055, -1.355) -- node[lab]{c} (n4) ;
\path[edge] (n3)  -- node[lab]{at} (n6) ;
\path[edge](n5.west |- 4.055, -0.855) -- node[lab]{at} (n6) ;
\path[edge] (n7)  -- node[lab]{at} (n6) ;
\path[edge](n8.east |- 1.055, -0.555) -- node[lab]{at} (n9) ;
\path[edge] (n10)  -- node[lab]{at} (n9) ;

\node[node, fill=none] at (6.95, -1.0)
{\begin{tabular}{@{}ll@{}}
Legenda: & \\
\textbf{FW} -- firewall & \textit{o} -- outer\\
\textbf{IF} -- interface & \textit{i} -- inner\\
\textbf{L} -- location & \textit{s} -- safe\\
\textbf{P} -- packet & \textit{u} -- unsafe
\vspace{-2pt}
\end{tabular}};
\userdefinedmacro
\end{tikzpicture}
\renewcommand{\userdefinedmacro}{\relax}
\caption{Concrete graph representing a possible state of a network with a
firewall.}
\flabel{ex-host-firewall}
\end{figure}
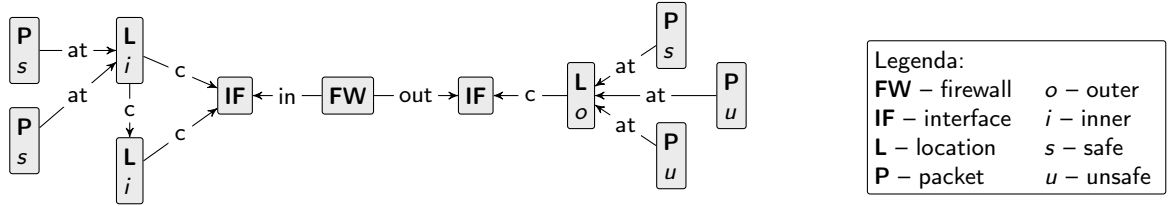

\fref{rules} shows four transformation rules of the grammar. The rules are in
\GROOVE single-graph notation, which uses colours and line formats to
distinguish rule elements. Black (continuous thin) elements are matched and kept
by rule application, blue elements (dashed thin) are matched and deleted, and
green (continuous bold) elements are created\footnote{In the standard two-graph
notation, the LHS is formed by black and blue elements and RHS by black and
green elements.}. \frefs{rules-a}{rules-b} show the rules for packet creation. A
safe packet can be created at any location, whereas an unsafe packet can only be
created at outer locations. Infinite behaviour stems from these two rules; since
they are always enabled, an infinite number of packets can be created.
\fref{rules-c} depicts a rule for packet transfer between locations. Since all
locations on each side of the firewall are of the same type, there is no need to
distinguish between safe and unsafe packets. A dual rule (not shown here)
transfers packets on the reverse direction of the connection edge, thus making
the connection bi-directional. The rule in \fref{rules-d} shows the firewall
filter, that only allows safe packets to reach inner locations.

\begin{figure}[t]
\centering
\subfigure[Safe packet creation]{\flabel{rules-a}
%
\begin{tikzpicture}[
scale=\tikzscale]
\node[node] (n0)  at (1.655, -1.855) {\ml{\textbf{L}}};
\node[newnode] (n1)  at (0.845, -1.815) {\ml{\textbf{P}\\\textit{s}}};
\path[newedge](n1.east |- 1.655, -1.855) -- node[newlab]{at} (n0) ;
\userdefinedmacro
\end{tikzpicture}
\renewcommand{\userdefinedmacro}{\relax}}
\hspace{2.5ex}
\subfigure[Unsafe packet creation]{\flabel{rules-b}
%
\begin{tikzpicture}[
scale=\tikzscale]
\node[newnode] (n1)  at (2.455, -1.355) {\ml{\textbf{P}\\\textit{u}}};
\node[node] (n0)  at (1.655, -1.355) {\ml{\textbf{L}\\\textit{o}}};
\path[newedge](n1.west |- 1.655, -1.355) -- node[newlab]{at} (n0) ;
\userdefinedmacro
\end{tikzpicture}
\renewcommand{\userdefinedmacro}{\relax}}
\hspace{2.5ex}
\subfigure[Packet transfer between locations]{\flabel{rules-c}
%
\begin{tikzpicture}[
scale=\tikzscale]
\node[node] (n1)  at (1.555, -1.355) {\ml{\textbf{L}}};
\node[node] (n0)  at (2.555, -1.355) {\ml{\textbf{L}}};
\node[node] (n2)  at (2.055, -0.655) {\ml{\textbf{P}}};
\path[newedge] (n2)  -- node[newlab]{at} (n0) ;
\path[deledge] (n2)  -- node[dellab]{at} (n1) ;
\path[edge](n1.east |- 2.555, -1.355) -- node[lab]{c} (n0) ;
\userdefinedmacro
\end{tikzpicture}
\renewcommand{\userdefinedmacro}{\relax}}
\hspace{2.5ex}
\subfigure[Firewall filtering]{\flabel{rules-d}
%
\begin{tikzpicture}[
scale=\tikzscale]
\node[node] (n1)  at (3.255, -2.055) {\ml{\textbf{IF}}};
\node[node] (n0)  at (2.405, -2.055) {\ml{\textbf{FW}}};
\node[node] (n2)  at (1.555, -2.055) {\ml{\textbf{IF}}};
\node[node] (n4)  at (3.955, -2.055) {\ml{\textbf{L}}};
\node[node] (n5)  at (2.455, -1.455) {\ml{\textbf{P}\\\textit{s}}};
\node[node] (n3)  at (0.855, -2.055) {\ml{\textbf{L}}};
\path[edge](n0.east |- 3.255, -2.055) -- node[lab]{out} (n1) ;
\path[edge](n3.east |- 1.555, -2.055) -- node[lab]{c} (n2) ;
\path[deledge] (n5)  -- node[dellab]{at} (n4) ;
\path[edge](n0.west |- 1.555, -2.055) -- node[lab]{in} (n2) ;
\path[newedge] (n5)  -- node[newlab]{at} (n3) ;
\path[edge](n4.west |- 3.255, -2.055) -- node[lab]{c} (n1) ;
\userdefinedmacro
\end{tikzpicture}
\renewcommand{\userdefinedmacro}{\relax}}
\caption{Examples of transformation rules for the firewall grammar.}
\flabel{rules}
\end{figure}
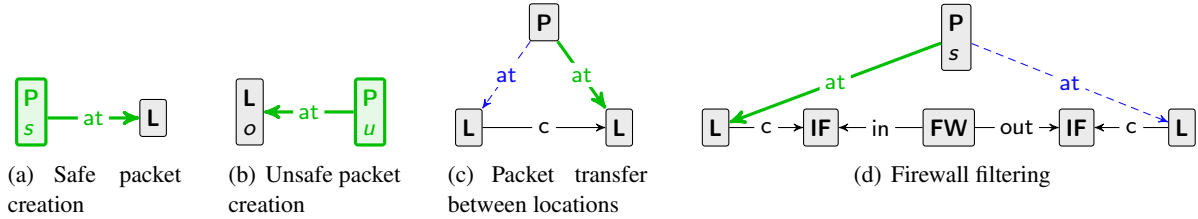

\subsection{Neighbourhood Abstraction}
\stlabel{neigh-abs}

Our notion of abstraction is based on neighbourhood similarity: nodes are
considered equivalent if they have the same labels and the same number of
incoming and outgoing edges. Graphs are abstracted by folding all equivalent
nodes into one, while keeping count of their original number up to some bound
of precision. Counting up to some bound is done using \emph{multiplicities}.

\subsubsection{Multiplicities}

We use $\omega$ to denote an upper bound on the set of natural numbers, \ie,
$\omega \notin \Nat$ and for all $k \in \Nat, k < \omega$. We write $\NatO$ as a
short-hand notation for $\Nat \setunion \set{\omega}$. We can then define simple
arithmetic operations over $\NatO$, such as \emph{addition} and
\emph{subtraction}. For example, given $i, j \in \NatO$,
\[
i + j = \left\{
\begin{array}{ll}
i + j & \textrm{ if } i, j \in \Nat,\\
\omega & \textrm{ otherwise}.
\end{array} \right.
\]
The symbol $+$ is overloaded: the one on the left represents addition over
$\NatO$ while the one on the right is the usual addition over $\Nat$. Note that
the first condition of the definition implies that $i + j < \omega$.

\begin{definition}[Multiplicity]
A \emph{multiplicity} is an element of the set $\calM = \set{\tup{i,j} \in (\Nat
\times \NatO) \st i \le j}$.
\defend
\end{definition}

Multiplicities are used to represent an interval of consecutive values taken
from $\NatO$, \ie, we write $\tup{i,j}$ as a compact representation for the set
$\set{k \in \NatO \st i \le k \le j}$. Given a multiplicity $\tup{i, j} \in
\calM$:
\begin{itemize}
\item if $i = j$, we call the multiplicity \emph{singleton} and we use a
short-hand notation by writing only the lower-bound $i$ in \textbf{bold}, \ie,
$\mathbf{i}$. The singleton multiplicity $\multone$ is called \emph{concrete};
and
\item if $j = \omega$, we use a short-hand notation by writing the lower-bound
$i$ in \textbf{bold}, super-scripted with $+$, \ie, $\mathbf{i}^{+}$.
\end{itemize}

Set $\calM$ is infinite, since $i$ and $j$ are taken from infinite sets. To
ensure finiteness, we need to define a bound of precision, which limits the
possible values of $i$ and $j$.

\begin{definition}[Bounded multiplicity]
A \emph{bounded multiplicity} is an element of set $\calMb \subset \calM$,
defined, for a given bound $\nbound \in \Nat$, as $\calMb = \{\tup{i, j}
\in \calM \st i \le \nbound + 1,\ j \in \set{0, \ldots, \nbound, \omega}\}$.
\defend
\end{definition}

The theory of neighbourhood abstraction presented in~\cite{BBKR08} is
parameterised with two multiplicity bounds, for node and edge counting. In
practice, these bounds are usually set to a low value, such as 1 or 2, since
they can greatly affect the size of the abstract state space. For the remainder
of this paper we consider both bounds to be 1 and we only work with bounded
multiplicities. This effectively limits the possible multiplicity values to six:
$\multz$, $\tup{0, 1}$, $\multzw$, $\multone$, $\multow$, and $\multtw$. Any
natural number can be projected to a bounded multiplicity by means of a
simple approximation function. For a given set $A$, we write $\setcard{A}$ to
denote the bounded multiplicity approximated from the cardinality of set $A$.

It is simple to define arithmetic operations over multiplicities based on the
operations over $\NatO$. In order to later define the state subsumption
relation (\stref{subsumption}) we need the concept of \emph{multiplicity
subsumption}, which amounts to interval inclusion. Given two bounded
multiplicities $\mu, \nu \in \calMb$, let $\mu = \tup{i, j}$ and $\nu = \tup{i',
j'}$. We say that $\mu$ is \emph{subsumed} by $\nu$ or that $\nu$
\emph{subsumes} $\mu$, denoted $\mu \subsump \nu$, if $i \ge i'$ and $j \le j'$.

\subsubsection{Neighbourhood Equivalence}

We begin this section introducing some additional notation. Given a graph $G
= \tup{N,E}$, a node $v \in N$, a binary label $l \in \LabB$, and a set of nodes
$C \subseteq N$, we consider the following sets of edges:
\begin{itemize}
\item $\oute(v, l, C) = \set{\tup{v, l, w} \in \EB \st w \in C}$, \ie, the set
of \emph{outgoing} $l$-edges from $v$ into nodes of $C$; and
\item $\ine(v, l, C) = \set{\tup{w, l, v} \in \EB \st w \in C}$, \ie, the set
of \emph{incoming} $l$-edges into $v$ from nodes of $C$.
\end{itemize}

Formally, to define neighbourhood similarity we create a \emph{neighbourhood
equivalence relation} $\neigheq$ over graph nodes.

\begin{definition}[Neighbourhood equivalence relation]
Given a graph $G = \tup{N,E}$, the \emph{neighbourhood equivalence relation}
$\neigheq$ over nodes of $G$ is defined for two radii (with $v, w \in N$):
\begin{itemize}
\item $v \neigheq_0 w$ if $\lab(v) = \lab(w)$; and
\item $v \neigheq_1 w$ if  $v \neigheq_0 w$, $\setcard{\oute(v, l, C)} =
\setcard{\oute(w, l, C)}$, and $\setcard{\ine(v, l, C)} = \setcard{\ine(w, l,
C)}$, for all binary labels $l \in \LabB$ and all sets of nodes $C \in
\seqclassz{N}$.
\defend
\end{itemize}
\end{definition}

From the definition, we see that two nodes are equivalent at radius 0 if they
have the same labels. Equivalence classes are then refined at radius 1, where
we look for the number of edges incoming from and outgoing to nodes of other
equivalence classes.

As with multiplicity bounds, the theory is also parameterised with an
abstraction radius. However, experiments with the prototype implementation
showed that increasing the radius above one is not feasible in practice. The
current abstraction implementation fixes the maximum radius to one.

\subsubsection{Shapes}

Our graph abstractions are called \emph{shapes}, following the term defined
in shape analysis~\cite{SRW02}. A shape is a graph with some additional
structure.

\begin{definition}[Shape]
A \emph{shape} is a tuple $S = \tup{G_S, {\simeq_S}, \multn_S, \multo_S,
\multi_S}$,
where:
\begin{itemize}
\item $G_S = \tup{N_S, E_S}$ is the underlying graph structure of the shape;
\item ${\simeq_S} \subseteq N_S \times N_S$ is a \emph{similarity relation}
over nodes of $S$;
\item $\multn_S \ftype {N_S}{\calM}$ is a \emph{node multiplicity} function,
which records how many concrete nodes were folded into an abstract node;
and
\item $\multo_S, \multi_S \ftype {(N_S \times \LabB \times
\seqclasseq{N_S}_S)}{\calM}$ are outgoing and incoming \emph{edge multiplicity}
functions, which record how many concrete edges with a certain label were folded
into an abstract edge.
\defend
\end{itemize}
\end{definition}

If a shape node $v$ has an associated concrete multiplicity, \ie,
$\multn_S(v) = \multone$, then $v$ is called \emph{concrete}. Nodes that are not
concrete are called \emph{collectors}.

A shape is constructed by abstracting a concrete host graph. Given a graph $G$,
we first compute the neighbourhood equivalence relation $\neigheq$ over $G$.
The graph component $G_S$ of the shape is constructed by folding the nodes of
each equivalence class of $\neigheq_1$, while recording the multiplicities of
these classes in the multiplicity maps of $S$. The similarity relation
${\simeq_S}$ is taken as $\neigheq_0$.\footnote{The definition of a shape is
``generic'' in the sense that any binary relation on nodes can be used as the
similarity relation~${\simeq}$. In this paper, however, we consider only shapes
where the relation ${\simeq}$ is taken as the neighbourhood equivalence
$\neigheq_0$.} A detailed explanation on shape construction is given
in~\cite{BBKR08}.

\begin{figure}[t]
\centering
%
\begin{tikzpicture}[
scale=\tikzscale]
\node[node] (n0)  at (2.550, -0.455) {\ml{\textbf{FW}}};
\node[node] (n1)  at (1.955, -1.155) {\ml{\textbf{IF}}};
\node[node] (n2)  at (3.155, -1.155) {\ml{\textbf{IF}}};
\node[node] (n3)  at (4.055, -1.155) {\ml{\textbf{L}\\\textit{o}}};
\node[node] (n4)  at (0.905, -1.355) {\ml{\textbf{L}\\\textit{i}}};
\node[node] (n5)  at (0.905, -0.555) {\ml{\textbf{L}\\\textit{i}}};
\node[node,ultra thick] (n6)  at (5.555, -1.155) {\ml{\textbf{P}\\\textit{u}}};
\node[node] (n7)  at (4.055, 0.305) {\ml{\textbf{P}\\\textit{s}}};
\node[node,ultra thick] (n8)  at (0.155, 0.305) {\ml{\textbf{P}\\\textit{s}}};

\draw[dashed] ($ (n0.north west) + (-0.1,0.1) $) rectangle
              ($ (n0.south east) + (0.1,-0.1) $);
\draw[dashed] ($ (n1.north west) + (-0.1,0.1) $) rectangle
              ($ (n2.south east) + (0.1,-0.1) $);
\draw[dashed] ($ (n3.north west) + (-0.1,0.1) $) rectangle
              ($ (n3.south east) + (0.1,-0.1) $);
\draw[dashed] ($ (n5.north west) + (-0.1,0.1) $) rectangle
              ($ (n4.south east) + (0.1,-0.1) $);
\draw[dashed] ($ (n6.north west) + (-0.1,0.1) $) rectangle
              ($ (n6.south east) + (0.1,-0.1) $);
\draw[dashed] ($ (n8.north west) + (-0.1,0.1) $) rectangle
              ($ (n7.south east) + (0.1,-0.1) $);

\node[circle,inner sep=0pt,minimum size=0pt,fill=none] (conn) at (1.5,
-1.175){};

\path[edge] (n0) -- node[lab]{in} (n1);
\path[edge] (n0) -- node[lab]{out} (n2);
\path[edge] (n3) -- node[lab]{c} (n2);
\path[edge,-] (n4) -- node[lab]{c} (conn);
\path[edge,-] (n5) -- node[lab]{c} (conn);
\path[edge] (conn) -- node[lab, above, pos=0.4, inner sep=0pt, font=\scriptsize]
{$2^+$} (n1);
\path[edge] (n5) -- node[lab]{c} (n4);
\path[edge] (n6) -- node[lab]{at} node[lab, pos=0.8, above, font=\scriptsize]{$2^+$} (n3);
\path[edge] (n7) -- node[lab]{at} (n3);
\path[edge] (n8) -- node[lab]{at} node[lab, pos=0.70, below,
font=\scriptsize,inner sep=0pt]{$2^+$} (n5);

\userdefinedmacro
\end{tikzpicture}
\renewcommand{\userdefinedmacro}{\relax}
\caption{Shape obtained when abstracting the graph of \fref{ex-host-firewall}.
Node multiplicities are indicated by thickness: bold nodes have multiplicity
$\multtw$; thin nodes have multiplicity $\multone$. Edge groups with
multiplicity $\multtw$ are explicitly shown; groups without values in the
figure have multiplicity $\multone$. Edge multiplicities close to the edge
arrow indicate incoming multiplicities from the opposite equivalence class.}
\flabel{ex-shape}
\end{figure}
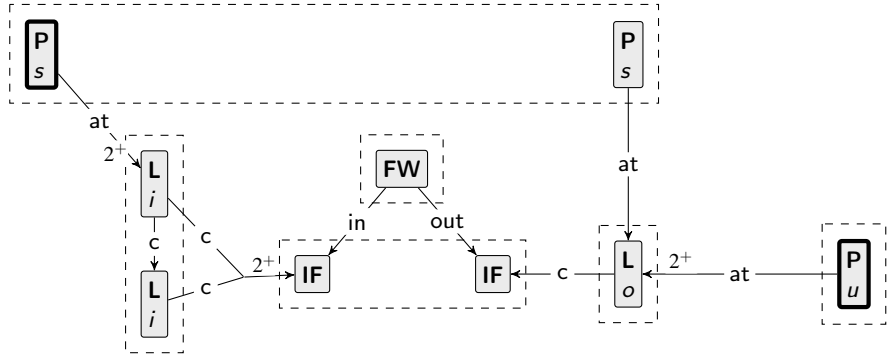

\fref{ex-shape} shows an example of a shape. The graph structure is drawn as
usual. The similarity relation {$\simeq_S$} is indicated with dashed boxes. Node
multiplicities are represented by line thickness: bold nodes have multiplicity
$\multtw$ and thin nodes have multiplicity $\multone$. Groups of edges with
incoming multiplicity $\multtw$ are explicitly identified; the remainder edge
multiplicities are all equal to $\multone$ and are not shown.
The shape in \fref{ex-shape} is an abstraction of the graph from
\fref{ex-host-firewall}, where the safe packets at the top inner location and
the unsafe packets at the outer location of the graph were collapsed into
collectors nodes of the shape (since the multiplicity bounds are one, any
natural greater than one is mapped to $\multtw$ in the abstract
domain\footnote{This loss of precision is intended, otherwise the abstraction
would not be finite.}). An important point to note is that the shape in
\fref{ex-shape} serves as an abstract representative not only for the graph in
\fref{ex-host-firewall} but also for any graph with \emph{two or more} packets
of the correct kind at the corresponding locations of the shape. Given a shape
$S$, we write $\concr(S)$ to indicate the (possibly infinite) set of
\emph{concretisations} of $S$, \ie, the set of graphs that can be abstracted to
$S$.


\section{Subsumption for State Space Reduction}
\stlabel{subsumption}

In this section we first present the general algorithm for abstract state space
traversal and we show the current duplicate detection mechanism used in \GROOVE,
based on graph certificates and graph isomorphism checks. We then proceed to
explain the new method of subsumption collapsing for abstract states and we
give a modified version for the traversal algorithm. This subsumption relation,
along with the experimental results given in \stref{experiments}, constitute
the new contributions of this paper.

\subsection{Abstract State Space Traversal}

\lstref{main-alg} gives the pseudo-code for exploring the abstract state space.
$Q$ is the set of all shapes and $F$ the set of fresh, yet to be explored
shapes; $R$ is the set of rules, $G$ the start graph, and $P$ is the set of
rule applications that were computed during exploration.
\begin{lstlisting}[caption=Algorithm for abstract state space exploration., label=\lstlabel{main-alg}]
let `$S \::=\: \abstr(G),\enspace Q \::=\: \emptyset, \enspace P \::=\: \emptyset, \enspace F \::=\: \set{S}$`
while `$F \neq \emptyset$`
do choose `$S\in F$ \quad` // which `$S$` is selected depends on the exploration strategy
   let `$F \::=\: F \setminus \set S$`
   for `$r \in R,\enspace m \in \prematch(r,S), \enspace S'\in \materialise(m,S)$`
   do let `$T \::=\: \normalise(\apply(r,m,S'))$`
      if `$\isfresh(T,Q,F)$ \quad` // if `$T \notin Q$`
      then let `$Q \::=\: Q\cup\set{T}, \enspace F \::=\: F\cup\set{T}$`
      fi
      let `$P \::=\: P\cup\set{S \transforms{r,m} T}$`
   od
od
\end{lstlisting}
The main phases in this algorithm are:
\begin{itemize}
\item \abstr computes the shape of a graph, as explained in the previous
section.
\item \prematch computes non-injective morphisms of a rule $r$ into a shape $S$.
Such a morphism is not yet a match, because the images of $r$'s LHS may be
collector elements; in this case they have to be materialised.
\item \materialise creates concrete nodes and edges for the image of $r$ in $S$.
This is a non-deterministic step, as there may be options for choosing
multiplicities for the materialised elements.
\item \apply is rule application, which can be carried out as usual because the
rule now acts upon a concrete subgraph of $S'$. At this step, the match of the
rule is injective.
\item \normalise merges the transformed graph back into the rest of the shape;
it is thus similar to \abstr except that it acts upon a (partially materialised)
shape rather than a graph.
\end{itemize}
Each of these phases are explained in detail in~\cite{RZ10}. For the purposes
of this paper it suffices to focus on the operations at lines 3 and 7 of the
algorithm given in \lstref{main-alg}.

Line 3 deals with the policy for selecting a shape $S$ from the set $F$ of
shapes to be explored. We consider two policies, namely Breadth-First Search
(BFS) and Depth-First Search (DFS). When using BFS, $F$ is implemented as a
queue, whereas in DFS $F$ is a stack. We use the term \emph{exploration
strategies} to refer to these search policies. \stref{experiments} gives an
experimental comparison on the performance of these two strategies in abstract
state space exploration.

Line 7 handles the duplicate state detection mechanism. Procedure
$\isfresh(T,Q,F)$ is responsible for checking if shape $T$ (or an equivalent
canonical representative) is already in the set $Q$ of all explored shapes.
Set $Q$ can be quite large, so this check has to be implemented with care, since
it can greatly impact performance. This is even more important when working
with graph grammars; in particular, in \GROOVE, states are collapsed under an
isomorphic representative. Since graph isomorphism can be a rather expensive
check, it cannot be performed over all elements of $Q$.

\lstref{isfresh} gives the algorithm for procedure $\isfresh$, as originally
described in~\cite{Ren06}. This algorithm is based on \emph{graph certificates},
which basically correspond to a hashing method tuned for graphs. In addition
to being relatively inexpensive to compute, certificates are built in such a way
that graph isomorphism implies certificate equality. The converse, however, is
not true, since the certificate function may produce false positives. In the
algorithm of \lstref{isfresh}, certificates are used to filter elements of set
$Q$, thus producing a much smaller set $\bar Q \subseteq Q$, composed only of
graphs with the same certificate of $T$. We then proceed to check if there
exists $U \in \bar Q$ such that $T$ and $U$ are isomorphic (denoted $T \iso U$).
If no such $U$ is found then we can conclude that $T$ is fresh.

\begin{lstlisting}[caption=Algorithm for procedure $\isfresh(T\!{,}\ Q\!{,}\ F)$., label=\lstlabel{isfresh}]
let `$C \::=\: \cert(T)$, \enspace $\bar Q \::=\: \set{U \in Q \st C = \cert(U)}$`
for `$U \in \bar Q$`
do if `$T \iso U$ \quad` // if `$T$ and $U$` are isomorphic
   then return false
   fi
od
return true `\quad` // we checked all candidates but none are isomorphic to `$T$`, so `$T$` is fresh
\end{lstlisting}

The method just described works very well in practice for concrete state space
exploration and since shapes also have a graph structure, we can immediately
reuse the same algorithm of \lstref{isfresh} for abstract exploration. However,
shapes carry additional information that is not taken into account when using
only isomorphism checks. To use this additional information, the notion of
\emph{shape subsumption} was developed.

\subsection{Shape Subsumption}

The key insight behind the shape subsumption relation (denoted by the same
symbol $\subsump$ used for multiplicity subsumption) lies in the comparison
between the concretisations of shapes. Let $S$ and $T$ be two \emph{isomorphic}
shapes, with $\concr(S) \subseteq \concr(T)$. Since $T$ has more concretisations
than $S$, rule applications on $T$ capture more behaviour than rule applications
on $S$. In fact, all behaviour of $S$ is subsumed by the behaviour of $T$, and
therefore, in an abstract exploration we can discard $S$ and only explore~$T$.
More formally, if $S$ is subsumed by $T$ then for all $S \transforms{r} U \in P$
there exists $T \transforms{r} U' \in P$ such that $U \subsump U'$, for any $r
\in R$.

Subsumption is an asymmetric relation built upon isomorphism. Shape $S$ is
subsumed by shape $T$ if: (i) $S$ and $T$ are isomorphic, (ii) they have the
same node similarity relation, and (iii) all multiplicities in $S$ are subsumed
by the multiplicities in $T$. Formally, we have the following definition, where,
for $v \in N$, we write $\eqclass{v}{\simeq}$ to denote the equivalence class of
$v$ induced by $\simeq$, \ie, $\eqclass{v} {\simeq}= \set{w \in N \st v \simeq
w}$.

\begin{definition}[Shape subsumption]
Given two shapes $S$ and $T$, $S$ is subsumed by $T$, denoted $S \subsump T$,
if:
\begin{itemize}
\item there exists an isomorphism $\isomap : G_S \iso G_T$ between the graph
structures of the shapes;
\item for any $\tup{v, w} \in\ \simeq_S$, $\tup{\isomap(v), \isomap(w)} \in\
\simeq_T$;
\item $\multn_S(v) \subsump \multn_T(\isomap(v))$, for all $v \in N_S$; and
\item $\multo_S(v, l, \eqclass{w}{\simeq_S}) \subsump \multo_T(\isomap(v), l,
\eqclass{\isomap(w)}{\simeq_T})$ and $\multi_S(v, l, \eqclass{w}{\simeq_S})
\subsump \multi_T(\isomap(v), l, \eqclass{\isomap(w)}{\simeq_T}),\ $ for all
$\tup{v, l, w} \in E_S$.
\defend
\end{itemize}
\end{definition}

Note that $S \subsump T$ implies $\concr(S) \subseteq \concr(T)$. As a simple
example, take $S$ as a shape containing only one node of multiplicity $\multtw$
and no edges. Then take $T$ also as a shape with a single node, but with
multiplicity $\multow$. From the definition, we have that $S \subsump T$. Set
$\concr(S)$ contains graphs with two or more nodes, whereas set $\concr(T)$ has
one more element, namely the graph with just one node. Hence, $\concr(S)
\subseteq \concr(T)$.

Shapes that subsume one another, \ie, shapes that are isomorphic and have the
same multiplicities for all elements, are called \emph{strictly isomorphic}.
From the above it follows that strictly isomorphic shapes have the same
concretisations (see the accompanying technical report of~\cite{BBKR08} for the
proof).

\lstref{newisfresh} shows the modified algorithm for procedure $\isfresh$, with
shape subsumption checks incorporated. Given a new shape $T$ that must be
tested for freshness, we begin as before, by constructing set $\bar Q$,
consisting of the shapes in $Q$ with the same certificate as $T$. In addition,
we initialise an auxiliary set $B$, to store shapes from $\bar Q$ that were
identified as subsumed by $T$. Since subsumption is an asymmetric relation, we
must check it in both directions (lines 3 and 5 of the algorithm). Note,
however, that these two subsumption checks do not require two isomorphism
checks, since we can first look for an isomorphism between $T$ and $U$ (the
potentially most expensive operation) and then proceed to check both
subsumptions using the same isomorphism.

\begin{lstlisting}[caption=Algorithm for procedure $\isfresh(T\!{,}\ Q\!{,}\ F)$ with
subsumption check., label=\lstlabel{newisfresh}]
let `$C \::=\: \cert(T)$, \enspace $\bar Q \::=\: \set{U \in Q \st C = \cert(U)}$, \enspace $B \::=\: \emptyset$`
for `$U \in \bar Q$`
do if `$T \subsump U$ \quad` // if `$T$` is subsumed by `$U$`
   then return false
   else if `$U \subsump T$ \quad` // if `$U$` is subsumed by `$T$`
        then let `$B \::=\: B\cup\set{U}$ \quad` // mark `$U$` as subsumed
        fi
   fi
od
`$F \::=\: F \setminus B$ \quad` // remove the states marked as subsumed from the set of states to be explored
return true
\end{lstlisting}

An interesting aspect of the new $\isfresh$ procedure is that it can now modify
the set $F$ of shapes to be explored. If there exists $U \in \bar Q$ such that
$T \subsump U$, then $T$ is not fresh and can be discarded as before (lines 3
and 4 of \lstref{newisfresh}). However, if we discover that $U \subsump T$, then
not only we know that $T$ is fresh, but also that $U$ should not be explored,
since all its behaviour is subsumed by $T$. We then mark $U$ as subsumed by
adding it to set $B$ (line 6) and continue looking for other subsumed shapes in
$\bar Q$. At the end of the procedure we remove all shapes marked as subsumed
from $F$ (line 10), thus trimming the search space.

\section{Experiments and Results}
\stlabel{experiments}

The theory of neighbourhood abstraction ensures that the number of shapes for
any graph grammar is finite, and therefore that the abstract state space is also
finite~\cite{BBKR08}. However, the theoretical upper bound of the abstract
state space size is still quite large, meaning that in practice we have to
optimise the state space traversal in order to implement an efficient tool.
In~\cite{RZ10} we described the major aspects of the implementation of
neighbourhood abstraction in \GROOVE (though without shape subsumption), and
reported a few experiments. In this section we present more of
such experiments, that illustrate: (i) the upper bounds on abstract state space
sizes that are reachable in practice, (ii) the performance gains that are
obtained with the shape subsumption technique presented in \stref{subsumption},
and (iii) how different exploration strategies perform in the abstract setting.

For our experiments we collected 8 graph grammars, from various problem
domains. Some of these problems can also be solved with other abstractions, but
they are usually tuned with certain characteristics from the domain at
hand\footnote{Shape analysis, for example, is designed to work on heap pointer
structures, which are deterministic graphs. In our setting, this correspond to
having shapes with all outgoing edge multiplicities limited to $\multone$.}. Our
abstraction extension for \GROOVE, on the other hand, is generic, in the sense
that the concept of neighbourhood equivalence is always applicable, but perhaps
with varying performance. Here is a list with a short description of the
grammars used.
\begin{itemize}
\item {\bf linked-list}: a grammar modelling operations on a single-linked
list structure. Elements can always be appended at the end of the list, which
can thus grow unbounded.
\item {\bf circ-buf-0}: a grammar modelling a circular buffer structure with
an unbounded number of cells. Cell usage is marked with special labels, without
reference to stored objects.
\item {\bf circ-buf-1}: a variant of the circular buffer where the stored
objects are explicitly represented.
\item {\bf euler-0}: a grammar that can construct Euler's walks of arbitrary
size. Adapted from the classical K\"onigsberg bridges problem from graph theory.
\item {\bf euler-1}: a variant of the Euler grammar without explicitly
representing connecting bridges.
\item {\bf firewall-[2-6]}: the grammar of our firewall example. Network
structure is fixed, while packages are collapsed by the abstraction. Instances
vary on the number of locations: from 2 to 6.
\item {\bf firewall-6-F}: variant of the firewall grammar with a network of six
fully connected locations.
\item {\bf car-platoon}: grammar simulating a wireless communication protocol
between cars, for establishing platoons in highways. Cars can enter and leave a
platoon at any time, which leads to an exponential growth on the number of
possible configurations.
\end{itemize}

\tref{size} gives all the figures on state space sizes for the grammars listed
above. Numbers for the BFS and DFS exploration strategies are grouped per
grammar, to ease the comparison between the two. State space sizes are divided
in two groups of values: the number of explored states, \ie, the number of
shapes produced, and the number of transitions between states, \ie, the count of
rule applications. State and transition counts in \tref{size} are broken down in
five and three types, respectively.

\begin{table}[t]
\centering
\caption{State space sizes for the explorations performed with different graph
grammars.}
\tlabel{size}
\scriptsize
\begin{tabular}{c|c|rrrrr|rrr}
\hline
\bf \multirow{2}{*}{Grammar} & \bf \multirow{2}{*}{Strat.} &
\multicolumn{5}{|c|}{\bf States} & \multicolumn{3}{|c}{\bf Transitions} \\
& & \bf Maximum & \bf Generated & \bf Subsumed & \bf Relevant & \bf Discarded
& \bf Maximum & \bf Generated & \bf Relevant \\
\hline
\multirow{2}{*}{linked-list} &
BFS & \multirow{2}{*}{9} & 9 & 3 & 6 & 0 & \multirow{2}{*}{17} & 17 & 11 \\
& DFS & & 8 & 2 & 6 & 1 & & 14 & 11 \\
\hline
\multirow{2}{*}{circ-buf-0} &
BFS & \multirow{2}{*}{54} & 31 &  5 & 26 & 3 & \multirow{2}{*}{130} & 65 & 59 \\
& DFS & & 39 & 13 & 26 & 2 & & 88 & 59 \\
\hline
\multirow{2}{*}{circ-buf-1} &
BFS & \multirow{2}{*}{57} & 33 & 16 & 17 & 0 & \multirow{2}{*}{182} & 100 & 40
\\
& DFS & & 30 & 13 & 17 & 2 & & 89 & 40 \\
\hline
\multirow{2}{*}{euler-0} &
BFS & \multirow{2}{*}{878} & 248 & 96 & 152 & 54 & \multirow{2}{*}{10,448} &
1,356 & 584 \\
& DFS & & 213 & 61 & 152 & 28 & & 1,286 & 618 \\
\hline
\multirow{2}{*}{euler-1} &
BFS & \multirow{2}{*}{14} & 14 & 4 & 10 & 0 & \multirow{2}{*}{23} & 23 & 15 \\
& DFS & & 13 & 3 & 10 & 2 & & 19 & 15 \\
\hline
\multirow{2}{*}{firewall-2} &
BFS & \multirow{2}{*}{125} & 98 & 90 & 8 & 36 & \multirow{2}{*}{875} & 409 & 37
\\
& DFS & & 50 & 42 & 8 & 15 & & 207 & 37 \\
\hline
\multirow{2}{*}{firewall-3} &
BFS & \multirow{2}{*}{1,625} & 991 & 971 & 20 & 549 & \multirow{2}{*}{19,825} &
5,021 & 121 \\
& DFS & & 232 & 212 & 20 & 102 & & 1,314 & 121 \\
\hline
\multirow{2}{*}{firewall-4} &
BFS & \multirow{2}{*}{4,875} & 2,356 & 2,326 & 30 & 1,487 &
\multirow{2}{*}{83,850} & 13,577 & 203 \\
& DFS & & 427 & 397 & 30 & 212 & & 2,959 & 203 \\
\hline
\multirow{2}{*}{firewall-5} &
BFS & \multirow{2}{*}{} & 14,878 & 14,818 & 60 & 10,549 & \multirow{2}{*}{} &
93,549 & 459 \\
& DFS & & 1,201 & 1,141 & 60 & 654 & & 10,421 & 459 \\
\hline
\multirow{2}{*}{firewall-6} &
BFS & \multirow{2}{*}{} & 25,251 & 25,171 & 80 & 18,373 & \multirow{2}{*}{} &
187,126 & 643 \\
& DFS & & 1,783 & 1,703 & 80 & 1,007 & & 18,485 & 643 \\
\hline
\multirow{2}{*}{firewall-6-F} &
BFS & \multirow{2}{*}{} & 183,478 & 182,966 & 512 & 147,028 & \multirow{2}{*}{}
& 1,409,451 & 8,711 \\
& DFS & & 5,930 & 5,418 & 512 & 3,003 & & 93,087 & 8,711 \\
\hline
car-platoon & DFS & \multicolumn{8}{|c}{Out of memory after exploring 445,439
states and 8,484,600 transitions} \\
\hline
\end{tabular}
\end{table}

The following five types of state count are given in columns 3-7 of
\tref{size}.
\begin{itemize}
\item {\bf Maximum} is the upper bound on the number of abstract states of the
grammar. This number is obtained by exploring the state space without shape
subsumption, \ie, by running the exploration algorithm of \lstref{main-alg}
with the original $\isfresh$ procedure of \lstref{isfresh}. Thus, these are the
figures that would have been reported by the prior implementation
of~\cite{RZ10}. Empty entries in
this column for the larger cases of the firewall grammar indicate that the
upper bound could not be computed: these runs timed out after several hours of
execution, due to state space explosion. Note that we give only one maximum
state count for both BFS and DFS strategies. The reason is that all states are
explored when subsumption is off, and therefore the maximum state count is
the same, regardless of the strategy used. The upper bound provided by this
column gives an interesting basis of comparison when analysing the reduction
obtained with subsumption.
\item {\bf Generated} is the number of states produced during exploration using
shape subsumption, \ie, the exploration algorithm of \lstref{main-alg} was run 
with the new $\isfresh$ procedure of \lstref{newisfresh}. Numbers in this
column correspond to the size of set $Q$, \ie, the number of states that were
added to the set of all explored states (line 8 in \lstref{main-alg}). When
comparing the number of generated states against the maximum upper bound we can
see the reduction given by subsumption. Take, for example, the
\emph{firewall-4} line, where the number of generated states using DFS with
subsumption is an order of magnitude smaller than the maximum upper bound
obtained without subsumption. The gain provided by subsumption can also be seen
for the larger cases of the firewall grammar: runs that timed-out without
subsumption can now be finished when we turn it on.
\item {\bf Subsumed} is the number of states generated that were later marked
as subsumed by another state. They correspond to states that are added to set
$B$ at line 6 of \lstref{newisfresh}.
\item {\bf Relevant} is the number of states that were never marked as subsumed
during the exploration. This column corresponds to the {\bf Generated} column
minus the {\bf Subsumed} one. The closer the number of generated states gets to
the relevant state count the better. A perfect exploration method would generate
only the relevant abstract states, since they are sufficient to cover all
concrete behaviour.
\item {\bf Discarded} is the number of states that were marked as subsumed and
were removed from the set of states to be explored. This number corresponds to
the sum of all states removed from $F$ at line 10 of \lstref{newisfresh}.
\end{itemize}
The remainder columns of \tref{size} give the number of transitions outgoing
from the states in the associated state count columns.

When comparing the figures in \tref{size} for BFS and DFS exploration, it is
clear that DFS gives a much better performance. The DFS generated state count is
smaller than the BFS count in all but one test (\emph{circ-buf-0}), and as we
move to grammars with larger state spaces the advantage increases greatly, until
reaching two orders of magnitude for the \emph{firewall-6-F} case. The reason
for this performance difference between BFS and DFS is simple. Usually, the more
a shape is transformed by subsequent rule applications the more abstract it
becomes, until it reaches a fix-point, \ie, further rule applications yield the
same shape again. These more abstract shapes capture more concrete behaviour
and thus can subsume other shapes in the state space. As a rule-of-thumb, more
abstract shapes are more likely to be part of the set of relevant states, and
since they are only discovered after a succession of rule applications, these
relevant states are deeper in the state space. Therefore, DFS is more likely to
reach these states first. This fact can be seen from the numbers in the {\bf
Discarded} column: BFS generates a lot of states that are later discarded. This
is wasted effort: in BFS a state is produced and added to sets $Q$ and $F$ but
it is very likely that later it is going to be removed from $F$ (while remaining
in $Q$). On the other hand, since DFS already found more abstract shapes, it is
more probable that the search will immediately throw a new state away, without
storing it on $Q$, since the new state will be subsumed by some other state
already in $Q$.

From the discussion above, one may wonder why shapes that were marked as
subsumed are kept in set $Q$. The reason is that removing states from $Q$ could
leave ``dangling'' transitions in the set of generated transitions $P$. A
possible solution could be the following. For $S \in Q$ and $T$ fresh, if $S
\subsump T$ then we should take all transitions in $P$ with $S$ as a source or
target and replace $S$ by $T$. This on-the-fly state space collapsing under
subsumption is not provided by the current implementation, but the tool offers
a simpler option: \emph{reachability mode}. In this mode we are only interested
in the shapes that are reachable in the abstract state space, and thus there is
no need to store the transitions, \ie, set $P$ is kept empty. In this case there
is no danger of having ``dangling'' transitions and we can remove subsumed
shapes from $Q$, thus decreasing memory usage. Reachability mode was a late
addition to the implementation and as such its experimental analysis is left as
future work.

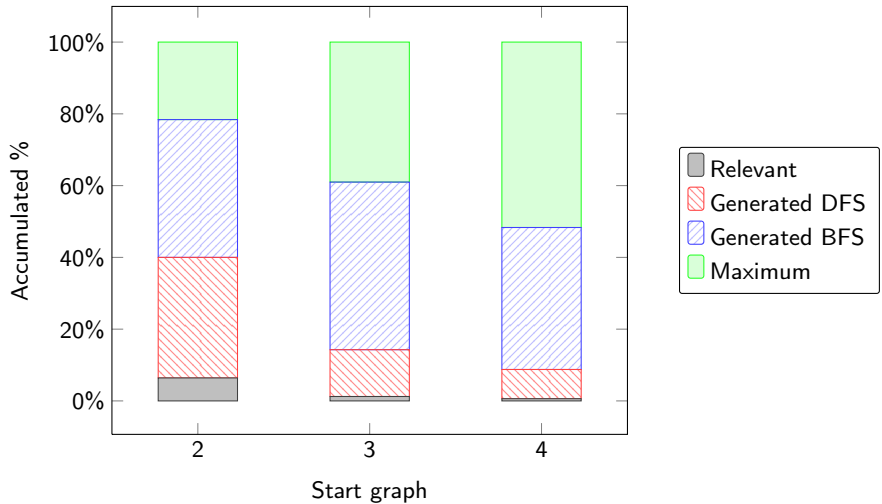
\begin{figure}[t]
\centering
\begin{tikzpicture}
\begin{axis}[
  ybar stacked,
  symbolic x coords={2, 3, 4, 5, 6, 6-F},
  yticklabels={0, 0\%, 20\%, 40\%, 60\%, 80\%, 100\%},
  xtick=data,
  bar width=30pt,
  enlarge x limits=0.25,
  xlabel=Start graph,
  ylabel=Accumulated \%,
  legend style={at={(1.1,0.5)},anchor=west, cells={anchor=west}},
]

\addplot[ybar,draw=black!75,fill=black!25] coordinates {
  (2, 0.064000) (3, 0.012308) (4, 0.006154)
};
\addplot[ybar,draw=red!75,pattern=north west lines,pattern color=red!45]
coordinates {
  (2, 0.336000) (3, 0.130462) (4, 0.081436)
};
\addplot[ybar,draw=blue!75,pattern=north east lines,pattern color=blue!25]
coordinates {
  (2, 0.384000) (3, 0.467077) (4, 0.395692)
};
\addplot[ybar,draw=green!85,fill=green!15] coordinates {
  (2, 0.216000) (3, 0.390154) (4, 0.516718)
};

\legend{Relevant, Generated DFS, Generated BFS, Maximum}
\end{axis}
\end{tikzpicture}
\vspace{-2ex}
\caption{Accumulated percentages of the number of states relative to the
maximum state space size, for the firewall grammar instances with a known
maximum state count.}
\flabel{state-chart}
\end{figure}

\fref{state-chart} gives a visual aid for the comparison between BFS and
DFS, for the firewall grammar instances with a known maximum state count. The
figure shows a stacked bar chart with the accumulated percentages of the number
of states relative to the maximum state space size. From this chart we see that
the percentage of states generated with DFS decreases as the start graph size
increases. On the other hand, the percentage of states generated with BFS
remains roughly the same, at around 40\% the maximum (this can be seen from the
interval sizes of the generated BFS bars in \fref{state-chart}).

Other metrics that must be analysed in a performance evaluation are running time
and memory consumption. Results for these measurements are given in \tref{time},
for runs with and without state subsumption checks. The experiments were
performed in a machine with a Intel Xeon X5365 CPU running at 3 GHz and a total
32 GB of RAM. Blank entries indicate timed-out executions. From the numbers in
\tref{time} we see the performance improvement given by subsumption: running
times for the firewall grammar are two orders of magnitude smaller when
subsumption is used, and memory consumption is also reduced. When comparing the
running times for BFS and DFS, we see that both strategies have a similar
performance when subsumption is not used but when it is turned on, DFS is far
more efficient than BFS, both in execution time and memory consumption. This
performance figures are directly related to the number of states generated by
each strategy: DFS produces far fewer states than BFS, which translates to a
large performance gain. This can be confirmed visually with the chart in
\fref{time-chart}, where the running times for the firewall grammar are plotted
against start graph sizes. Clearly, DFS has a much more tamed growth (note that
the time axis is in a logarithmic scale).

\begin{table}[t]
\centering
\caption{Running time and memory consumption, with and without state subsumption
checks.}
\tlabel{time}
\scriptsize
\begin{tabular}{c|rr|rr|rr|rr}
\hline
\bf \multirow{3}{*}{Grammar} & \multicolumn{4}{|c|}{\bf Time (s)} &
\multicolumn{4}{|c}{\bf Memory (MB)} \\
& \multicolumn{2}{|c|}{\bf Subsump. OFF} &
\multicolumn{2}{|c|}{\bf Subsump. ON} &
\multicolumn{2}{|c|}{\bf Subsump. OFF} &
\multicolumn{2}{|c}{\bf Subsump. ON} \\
& \bf BFS & \bf DFS
& \bf BFS & \bf DFS
& \bf BFS & \bf DFS
& \bf BFS & \bf DFS \\
\hline
linked-list & $<$ 1 & $<$ 1 & $<$ 1 & $<$ 1 & $<$ 1 & $<$ 1 & $<$ 1 & $<$ 1 \\
circ-buf-0 & $<$ 1 & $<$ 1 & $<$ 1 & $<$ 1 & 2 & 2 & 2 & 2 \\
circ-buf-1 & $<$ 1 & $<$ 1 & $<$ 1 & $<$ 1 & 2 & 2 & 2 & 2 \\
euler-0 & 61 & 47 & 2 & 2 & 57 & 57 & 12 & 11 \\
euler-1 & $<$ 1 & $<$ 1 & $<$ 1 & $<$ 1 & $<$ 1 & $<$ 1 & $<$ 1 & $<$ 1 \\
firewall-2 & 2 & 2 & $<$ 1 & $<$ 1 & 6 & 6 & 4 & 2 \\
firewall-3 & 177 & 157 & 5 & 2 & 110 & 110 & 49 & 12 \\
firewall-4 & 4,448 & 3,824 & 16 & 3 & 432 & 432 & 136 & 25 \\
firewall-5 & & & 347 & 10 & & & 1,054 & 85 \\
firewall-6 & & & 1,679 & 16 & & & 2,001 & 143 \\
firewall-6-F & & & 3,732 & 55 & & & 14,277 & 556 \\
\hline
\end{tabular}
\end{table}

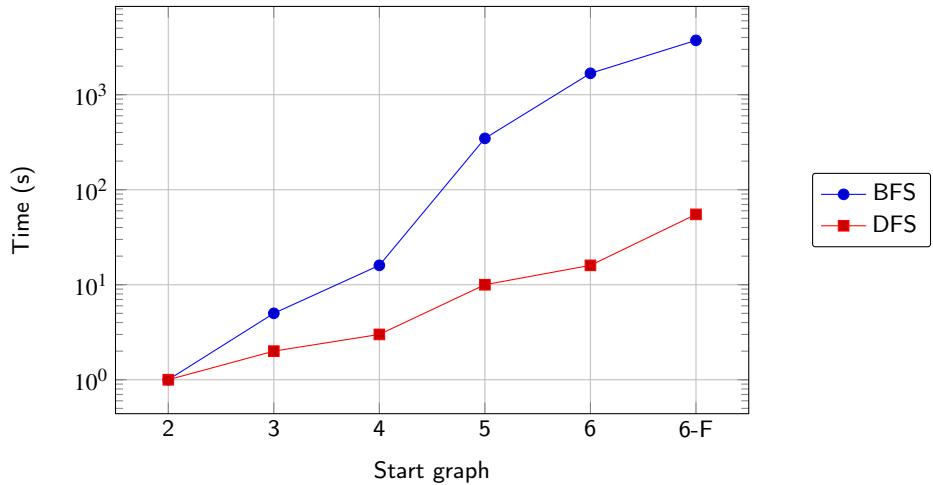
\begin{figure}[t]
\centering
\begin{tikzpicture}
\begin{semilogyaxis}[
  width=10cm,
  height=7cm,
  xlabel=Start graph,
  ylabel=Time (s),
  grid=major,
  symbolic x coords={2, 3, 4, 5, 6, 6-F},
  xtick=data,
  legend style={at={(1.1,0.5)}, anchor=west},
]

\addplot coordinates {
  (2, 1)
  (3, 5)
  (4, 16)
  (5, 347)
  (6, 1679)
  (6-F, 3732)
};
\addlegendentry{BFS}

\addplot coordinates {
  (2, 1)
  (3, 2)
  (4, 3)
  (5, 10)
  (6, 16)
  (6-F, 55)
};
\addlegendentry{DFS}

\end{semilogyaxis}
\end{tikzpicture}
\vspace{-2ex}
\caption{Running time (with subsumption on) versus start graph size for the
firewall grammar.}
\flabel{time-chart}
\end{figure}


\vspace{-1ex}
\section{Related Work}
\stlabel{relwork}

Abstraction is an essential ingredient in nearly all methods for system
analysis and verification and as such there is a vast body of work describing
the use of abstractions in different domains. In this section we give a
(non-exhaustive) discussion on related work that involves state space traversal
and graphs.

In~\cite{HMZM96}, Holte \emph{et al.} tackle the area of problem solving in
artificial intelligence, which boils down to finding the shortest path between
a start and a goal state.  There is a relation of opposition between solution
quality (the path length) and search effort (states traversed) and many
heuristics can be used to guide the search. The authors define a so-called
``explicit graph notation'', where the state space is represented by a labelled
transition system (LTS), and they proceed to define abstraction algorithms that
can be used to speed-up the search. One of such algorithms, called STAR, works
by building state classes that are connected up to a certain abstraction radius.
Despite having many similar concepts with our work, the abstractions used by
Holte are not state abstractions; they operate on the LTS level and not on the
state representation. Furthermore, the concrete state spaces considered
in~\cite{HMZM96} are always finite.

In~\cite{EJL06}, Edelkamp \emph{et al.}~consider the problem of partial
analysis/exploration of the state space of graph transformation systems. As
in the work of Holte \emph{et al.}, this amounts to a guided search over the
concrete state space where abstraction can be used as an heuristic. Properties
of interest for the analysis usually encompass existential checks for graph
structures; \eg, is a graph with a certain node and edge configuration
reachable from the start state? Any abstraction that preserves reachability of
the goal state in the abstract state space can be used to define an heuristic
for the guided search in the concrete level. Since our neighbourhood abstraction
preserves reachability, it could in principle be used as the abstraction
mechanism for an heuristic search. However, performance may be an issue, since
computing the transitions of an abstract state is a rather expensive operation.

Concerning the verification of infinite-state graph transformation systems,
K\"onig \emph{et al.}~have an extensive corpus of work, starting
with~\cite{BCK01}. Given a graph grammar their analysis technique extracts an
\emph{approximated unfolding}; a finite structure (called \emph{Petri graph})
that is composed of a hyper-graph and a Petri net. The Petri graph captures all
structure that can occur in the reachable graphs of the system, and dependencies
for rule applications are recorded by the Petri net transitions. The final Petri
graph obtained is an over-approximation that can be used to check safety
properties in the original system. If a spurious counter-example is introduced
by the over-approximation, the abstraction can be incrementally
refined~\cite{KK06}. These techniques are implemented in the tool {\sc augur}
which is now in its second version~\cite{AUGUR}. An experimental comparison
between this tool and our implementation is considered as future work.

\section{Conclusions and Future Work}
\stlabel{conclusion}

In this paper we present an abstraction technique for the exploration of
graph transformation systems with infinite state spaces. We explain the main
points of neighbourhood abstraction as implemented in \GROOVE and we propose a
new method for state collapsing, based on the concept of shape subsumption.
Experimental results show that subsumption gives a significant reduction on the
number of states that have to be explored, thus improving both the running time
and memory consumption of the tool. Furthermore, the experiments also show that
the choice of the exploration strategy has a heavy influence on performance,
with DFS giving much better results.

We see the results presented in this paper as an important achievement over the
original implementation of abstraction in \GROOVE. As any tool developer would
know, performance improvements in programs that deal with highly combinatorial
problems such as state space exploration usually involve a painstaking cycle of
refactorings, experimentation and fine-tuning. Our case was no different, where
the original abstraction code had to be rewritten from scratch in order to
accommodate shape subsumption. A further improvement over the code
from~\cite{RZ10} is that rules with NACs (negative application conditions) are
now also supported, which increases rule expressivity.

There are many directions where the current research/tool can be extended.
Aside from the usual points, such as additional experimentation with more test
cases and comparison with other tools, we consider the following items as future
work.
\begin{itemize}
%
\item {\bf Stronger notion of subsumption.} The subsumption relation presented
here depends on the existence of an isomorphism between two shapes. This
dependence can be weakened by requiring only the existence of an embedding
morphism between the shapes, which is not an isomorphism but instead an
injective sub-graph morphism, similar to a rule match. This weakening of the
subsumption pre-condition makes the relation stronger, and thus should lead to
further reductions of the state space. This new relation, however, requires
additional refactoring of the code, since we can no longer re-use the
isomorphism checking package from \GROOVE.
\item {\bf More expressive notions of abstraction.} While neighbourhood
abstraction can be used for many different classes of problems, it does not
fare very well when some structural properties should be preserved by the
abstraction. It cannot, for example, represent connectivity information between
nodes. When the abstraction does not limit the possible concrete structures
that can be generated, all cases have to be considered and this leads to a
blow-up in the abstract state space size that can cripple performance. We can
see this from the results for the \emph{car-platoon} grammar in \tref{size}: the
number of states in the state space is too large, and execution was aborted due
to an out-of-memory error. To tackle these problems, other notions of
abstraction are thus in order. We are currently working on the theory for a
pattern based abstraction, a method that will allow certain graph structures of
interest to be preserved in the abstract domain.
\end{itemize}

\bigskip

\noindent {\bf Availability.}
The current abstraction extension described in this paper is implemented in
\GROOVE version 4.4.6, available at \url{http://groove.cs.utwente.nl}. The
grammars for the experiments described in \stref{experiments} along with the
results obtained can also be downloaded at the same address.
%
%


\bibliographystyle{eptcs}
\bibliography{main}

\begin{thebibliography}{10}
\providecommand{\bibitemdeclare}[2]{}
\providecommand{\urlprefix}{Available at }
\providecommand{\url}[1]{\texttt{#1}}
\providecommand{\href}[2]{\texttt{#2}}
\providecommand{\urlalt}[2]{\href{#1}{#2}}
\providecommand{\doi}[1]{doi:\urlalt{http://dx.doi.org/#1}{#1}}
\providecommand{\bibinfo}[2]{#2}

\bibitemdeclare{book}{BK08}
\bibitem{BK08}
\bibinfo{author}{C.~Baier} \& \bibinfo{author}{J.~P. Katoen}
  (\bibinfo{year}{2008}): \emph{\bibinfo{title}{Principles of Model Checking}}.
\newblock \bibinfo{publisher}{MIT Press}, \bibinfo{address}{New York}.

\bibitemdeclare{inproceedings}{BCK01}
\bibitem{BCK01}
\bibinfo{author}{P.~Baldan}, \bibinfo{author}{A.~Corradini} \&
  \bibinfo{author}{B.~K{\"o}nig} (\bibinfo{year}{2001}):
  \emph{\bibinfo{title}{A Static Analysis Technique for Graph Transformation
  Systems}}.
\newblock In: {\sl \bibinfo{booktitle}{International Conference on Concurrency
  Theory (CONCUR)}}, {\sl \bibinfo{series}{LNCS}} \bibinfo{volume}{2154},
  \bibinfo{publisher}{Springer}, pp. \bibinfo{pages}{381--395}.
\newblock \urlprefix\url{http://dx.doi.org/10.1007/3-540-44685-0_26}.

\bibitemdeclare{inproceedings}{BBKR08}
\bibitem{BBKR08}
\bibinfo{author}{J.~Bauer}, \bibinfo{author}{I.~B. Boneva},
  \bibinfo{author}{M.~E. Kurban} \& \bibinfo{author}{A.~Rensink}
  (\bibinfo{year}{2008}): \emph{\bibinfo{title}{A Modal-Logic Based Graph
  Abstraction}}.
\newblock In \bibinfo{editor}{\bibinfo{editor}{Ehrig}} et~al.  \cite{ICGT2008},
  pp. \bibinfo{pages}{321--335}.
\newblock \urlprefix\url{http://dx.doi.org/10.1007/978-3-540-87405-8_22}.

\bibitemdeclare{inproceedings}{BW07}
\bibitem{BW07}
\bibinfo{author}{J.~Bauer} \& \bibinfo{author}{R.~Wilhelm}
  (\bibinfo{year}{2007}): \emph{\bibinfo{title}{Static Analysis of Dynamic
  Communication Systems by Partner Abstraction}}.
\newblock In: {\sl \bibinfo{booktitle}{Static Analysis Symposium (SAS)}}, {\sl
  \bibinfo{series}{LNCS}} \bibinfo{volume}{4634},
  \bibinfo{publisher}{Springer}, pp. \bibinfo{pages}{249--264}.
\newblock \urlprefix\url{http://dx.doi.org/10.1007/978-3-540-74061-2_16}.

\bibitemdeclare{inproceedings}{CC77}
\bibitem{CC77}
\bibinfo{author}{P.~Cousot} \& \bibinfo{author}{R.~Cousot}
  (\bibinfo{year}{1977}): \emph{\bibinfo{title}{Abstract Interpretation: A
  Unified Lattice Model for Static Analysis of Programs by Construction or
  Approximation of Fixpoints}}.
\newblock In: {\sl \bibinfo{booktitle}{Principles of Programming Languages
  (POPL)}}, \bibinfo{publisher}{ACM}, pp. \bibinfo{pages}{238--252}.
\newblock \urlprefix\url{http://doi.acm.org/10.1145/512950.512973}.

\bibitemdeclare{inproceedings}{EJL06}
\bibitem{EJL06}
\bibinfo{author}{S.~Edelkamp}, \bibinfo{author}{S.~Jabbar} \&
  \bibinfo{author}{A.~Lluch-Lafuente} (\bibinfo{year}{2006}):
  \emph{\bibinfo{title}{Heuristic Search for the Analysis of Graph Transition
  Systems}}.
\newblock In: {\sl \bibinfo{booktitle}{International Conference on Graph
  Transformations (ICGT)}}, {\sl \bibinfo{series}{LNCS}}
  \bibinfo{volume}{4178}, \bibinfo{publisher}{Springer}, pp.
  \bibinfo{pages}{414--429}.
\newblock \urlprefix\url{http://dx.doi.org/10.1007/11841883_29}.

\bibitemdeclare{book}{Handbook-GG-2}
\bibitem{Handbook-GG-2}
\bibinfo{editor}{H.~Ehrig}, \bibinfo{editor}{G.~Engels}, \bibinfo{editor}{H.-J.
  Kreowski} \& \bibinfo{editor}{G.~Rozenberg}, editors (\bibinfo{year}{1999}):
  \emph{\bibinfo{title}{Handbook of Graph Grammars and Computing by Graph
  Transformation: Applications, Languages, and Tools}}.
\newblock \bibinfo{publisher}{World Scientific Publishing Co.}

\bibitemdeclare{proceedings}{ICGT2008}
\bibitem{ICGT2008}
\bibinfo{editor}{H.~Ehrig}, \bibinfo{editor}{R.~Heckel},
  \bibinfo{editor}{G.~Rozenberg} \& \bibinfo{editor}{G.~Taentzer}, editors
  (\bibinfo{year}{2008}): \emph{\bibinfo{title}{International Conference on
  Graph Transformations (ICGT)}}. {\sl \bibinfo{series}{LNCS}}
  \bibinfo{volume}{5214}, \bibinfo{publisher}{Springer}.

\bibitemdeclare{article}{STTT}
\bibitem{STTT}
\bibinfo{author}{A.~Ghamarian}, \bibinfo{author}{M.~de~Mol},
  \bibinfo{author}{A.~Rensink}, \bibinfo{author}{E.~Zambon} \&
  \bibinfo{author}{M.~Zimakova} (\bibinfo{year}{2012}):
  \emph{\bibinfo{title}{Modelling and Analysis Using {\GROOVE}}}.
\newblock {\sl \bibinfo{journal}{International Journal on Software Tools for
  Technology Transfer (STTT)}} \bibinfo{volume}{14}(\bibinfo{number}{1}), pp.
  \bibinfo{pages}{15--40}.
\newblock \urlprefix\url{http://dx.doi.org/10.1007/s10009-011-0186-x}.

\bibitemdeclare{book}{God96}
\bibitem{God96}
\bibinfo{author}{P.~Godefroid} (\bibinfo{year}{1996}):
  \emph{\bibinfo{title}{Partial-Order Methods for the Verification of
  Concurrent Systems: An Approach to the State-Explosion Problem}}.
\newblock \bibinfo{publisher}{Springer Verlag}, \bibinfo{address}{New York}.

\bibitemdeclare{article}{HMZM96}
\bibitem{HMZM96}
\bibinfo{author}{R.~Holte}, \bibinfo{author}{T.~Mkadmi},
  \bibinfo{author}{R.~Zimmer} \& \bibinfo{author}{A.~MacDonald}
  (\bibinfo{year}{1996}): \emph{\bibinfo{title}{Speeding up Problem Solving by
  Abstraction: A Graph Oriented Approach}}.
\newblock {\sl \bibinfo{journal}{Artificial Intelligence}}
  \bibinfo{volume}{85}(\bibinfo{number}{1-2}), pp. \bibinfo{pages}{321--361}.
\newblock \urlprefix\url{http://dx.doi.org/10.1016/0004-3702(95)00111-5}.

\bibitemdeclare{inproceedings}{KK06}
\bibitem{KK06}
\bibinfo{author}{B.~K{\"o}nig} \& \bibinfo{author}{V.~Kozioura}
  (\bibinfo{year}{2006}): \emph{\bibinfo{title}{Counterexample-Guided
  Abstraction Refinement for the Analysis of Graph Transformation Systems}}.
\newblock In: {\sl \bibinfo{booktitle}{International Conference on Tools and
  Algorithms for the Construction and Analysis of Systems (TACAS)}}, {\sl
  \bibinfo{series}{LNCS}} \bibinfo{volume}{3920},
  \bibinfo{publisher}{Springer}, pp. \bibinfo{pages}{197--211}.
\newblock \urlprefix\url{http://dx.doi.org/10.1007/11691372_13}.

\bibitemdeclare{article}{AUGUR}
\bibitem{AUGUR}
\bibinfo{author}{B.~K{\"o}nig} \& \bibinfo{author}{V.~Kozioura}
  (\bibinfo{year}{2008}): \emph{\bibinfo{title}{Augur 2 - A New Version of a
  Tool for the Analysis of Graph Transformation Systems}}.
\newblock {\sl \bibinfo{journal}{Electronic Notes in Theoretical Computer
  Science (ENTCS)}} \bibinfo{volume}{211}, pp. \bibinfo{pages}{201--210}.
\newblock \urlprefix\url{http://dx.doi.org/10.1016/j.entcs.2008.04.042}.

\bibitemdeclare{inproceedings}{GROOVE}
\bibitem{GROOVE}
\bibinfo{author}{A.~Rensink} (\bibinfo{year}{2004}): \emph{\bibinfo{title}{The
  {\GROOVE} Simulator: A Tool for State Space Generation}}.
\newblock In: {\sl \bibinfo{booktitle}{Applications of Graph Transformations
  with Industrial Relevance (AGTIVE)}}, {\sl \bibinfo{series}{LNCS}}
  \bibinfo{volume}{3062}, \bibinfo{publisher}{Springer}, pp.
  \bibinfo{pages}{479--485}.
\newblock \urlprefix\url{http://dx.doi.org/10.1007/978-3-540-25959-6_40}.

\bibitemdeclare{inproceedings}{Ren04}
\bibitem{Ren04}
\bibinfo{author}{A.~Rensink} (\bibinfo{year}{2004}):
  \emph{\bibinfo{title}{Canonical Graph Shapes}}.
\newblock In: {\sl \bibinfo{booktitle}{European Symposium on Programming
  (ESOP)}}, {\sl \bibinfo{series}{LNCS}} \bibinfo{volume}{2986},
  \bibinfo{publisher}{Springer}, pp. \bibinfo{pages}{401--415}.
\newblock \urlprefix\url{http://dx.doi.org/10.1007/978-3-540-24725-8_28}.

\bibitemdeclare{inproceedings}{Ren06}
\bibitem{Ren06}
\bibinfo{author}{A.~Rensink} (\bibinfo{year}{2006}):
  \emph{\bibinfo{title}{Isomorphism Checking in \GROOVE}}.
\newblock In: {\sl \bibinfo{booktitle}{International Workshop on Graph-Based
  Tools (GraBaTs)}}, {\sl \bibinfo{series}{Electronic Communications of the
  EASST}}~\bibinfo{volume}{1}, \bibinfo{publisher}{European Association of
  Software Science and Technology}.
\newblock
  \urlprefix\url{http://eceasst.cs.tu-berlin.de/index.php/eceasst/article/view/77}.

\bibitemdeclare{inproceedings}{RD06}
\bibitem{RD06}
\bibinfo{author}{A.~Rensink} \& \bibinfo{author}{D.~Distefano}
  (\bibinfo{year}{2006}): \emph{\bibinfo{title}{Abstract Graph
  Transformation}}.
\newblock In: {\sl \bibinfo{booktitle}{Workshop on Software Verification and
  Validation (SVV)}}, {\sl \bibinfo{series}{Electronic Notes in Theoretical
  Computer Science (ENTCS)}} \bibinfo{volume}{157}, pp.
  \bibinfo{pages}{39--59}.
\newblock \urlprefix\url{http://dx.doi.org/10.1016/j.entcs.2006.01.022}.

\bibitemdeclare{inproceedings}{RZ10}
\bibitem{RZ10}
\bibinfo{author}{A.~{Rensink}} \& \bibinfo{author}{E.~{Zambon}}
  (\bibinfo{year}{2010}): \emph{\bibinfo{title}{Neighbourhood Abstraction in
  \GROOVE}}.
\newblock In: {\sl \bibinfo{booktitle}{International Workshop on Graph-Based
  Tools (GraBaTs)}}, {\sl \bibinfo{series}{Electronic Communications of the
  EASST}}~\bibinfo{volume}{32}, \bibinfo{publisher}{European Association of
  Software Science and Technology}.
\newblock
  \urlprefix\url{http://journal.ub.tu-berlin.de/index.php/eceasst/article/view/501}.

\bibitemdeclare{inproceedings}{RN08}
\bibitem{RN08}
\bibinfo{author}{S.~Rieger} \& \bibinfo{author}{T.~Noll}
  (\bibinfo{year}{2008}): \emph{\bibinfo{title}{Abstracting Complex Data
  Structures by Hyperedge Replacement}}.
\newblock In \bibinfo{editor}{\bibinfo{editor}{Ehrig}} et~al.  \cite{ICGT2008},
  pp. \bibinfo{pages}{69--83}.
\newblock \urlprefix\url{http://dx.doi.org/10.1007/978-3-540-87405-8_6}.

\bibitemdeclare{article}{SRW02}
\bibitem{SRW02}
\bibinfo{author}{S.~Sagiv}, \bibinfo{author}{T.~W. Reps} \&
  \bibinfo{author}{R.~Wilhelm} (\bibinfo{year}{2002}):
  \emph{\bibinfo{title}{Parametric Shape Analysis via 3-valued Logic}}.
\newblock {\sl \bibinfo{journal}{Transactions on Programming Languages and
  Systems (ToPLaS)}} \bibinfo{volume}{24}(\bibinfo{number}{3}), pp.
  \bibinfo{pages}{217--298}.
\newblock \urlprefix\url{http://doi.acm.org/10.1145/514188.514190}.

\bibitemdeclare{inproceedings}{SWJ08}
\bibitem{SWJ08}
\bibinfo{author}{M.~Saksena}, \bibinfo{author}{O.~Wibling} \&
  \bibinfo{author}{B.~Jonsson} (\bibinfo{year}{2008}):
  \emph{\bibinfo{title}{Graph Grammar Modeling and Verification of Ad Hoc
  Routing Protocols}}.
\newblock In: {\sl \bibinfo{booktitle}{International Conference on Tools and
  Algorithms for the Construction and Analysis of Systems (TACAS)}}, {\sl
  \bibinfo{series}{LNCS}} \bibinfo{volume}{4963},
  \bibinfo{publisher}{Springer}, pp. \bibinfo{pages}{18--32}.
\newblock \urlprefix\url{http://dx.doi.org/10.1007/978-3-540-78800-3_3}.

\end{thebibliography}

\end{document}